\documentclass[12pt]{iopart}
\usepackage{epsfig}
\usepackage{epsf}

\def\slash#1{\setbox0=\hbox{$#1$}
   \dimen0=\wd0 \setbox1=\hbox{/} \dimen1=\wd1
   \ifdim\dimen0>\dimen1 \rlap{\hbox to \dimen0{\hfil/\hfil}} #1
   \else  \rlap{\hbox to \dimen1{\hfil$#1$\hfil}} / \fi}

\begin{document}

\title{Correlation equalities and upper bounds for the transverse Ising model}

\author{F C S\'{a} Barreto and A L Mota}
\address{Departamento de Ci\^{e}ncias Naturais, Universidade Federal de S\~{a}o Jo\~{a}o del Rei,
C.P. 110,  CEP 36301-160, S\~ao Jo\~ao del Rei, Brazil}
\eads{\mailto{fcsabarreto@gmail.com},\mailto{ motaal@ufsj.edu.br}}

\begin{abstract}
Starting from an exact formal identity for the two-state transverse Ising model and using correlation inequalities rigorous upper bounds for the critical temperature and the critical transverse field are obtained which improve effective results.
\end{abstract}

\maketitle

\section{Introduction}

The transverse Ising model (TIM) is described by a two-state Ising Hamiltonian with a term representing a field transverse to the spins,

\begin{equation}
H = - \sum_{ij} J_{ij}S_i^zS_j^z - \Omega \sum_i S_i^x, \label{eq1}
\end{equation}   
where $J_{ij}>0$, $\Omega$ is the transverse field, $S_i^z$ and $S_i^x$ are Pauli spin-$\frac{1}{2}$ operators and the first sum is over the nearest neighbors spins on the lattice. 
The model has been firstly applied to describe the phase transitions and the properties of hydrogen bonded ferroelectrics \cite{deGennes1963,Blinc1972} and magnetic ordered materials \cite{Wang1968}. This model in one dimension has no phase transition at finite temperatures; however, at zero temperature it is ordered up to the critical value of the transverse field. 
The model has been solved exactly in one dimension \cite{Pfeuty1970, Katsura1968, Suzuki1971}. In high dimensions there are approximations for low-temperatures or high-temperatures regions \cite{Elliott1971, Pfeuty1971}. All other calculations are based on mean field type approximations. An effective field theory has been presented which improve over mean field results \cite{SaBarreto1981}. Since then many results have been obtained based on the effective field theory. More recently the model has been used to study the phase diagrams of nanowires systems \cite{Kaneyoshi2010} and magnetization of nanoparticles \cite{Kaneyoshi2009} . The objective of this paper is to present rigorous upper bounds for the critical couplings. We will apply the results for the d=2 square lattice and the d=3 cubic lattice.

\section{Generalized Callen's identity for the transverse Ising model}
 
In this section, we describe the methodology used by S\'{a} Barreto et al \cite{SaBarreto1981} to derive an identity for the two-spin correlation function of the transverse Ising model. The procedure used in this deduction was presented in reference \cite{SaBarreto1981} to obtain an exact relation for the order parameter $<$$S_i^z$$>$ which generalizes Callen's identity \cite{Callen1963}.
The longitudinal two-spin correlation function $<$$S_l^zS_i^z$$>$ can be calculated from
\begin{equation}
<S_l^zS_i^z> = \frac{ Tr(e^{-\beta H} S_l^zS_i^z)}{Tr(e^{-\beta H})} \label{eq2}     
\end{equation}  
where H is given by (\ref{eq1}). The Hamiltonian can be separated into two parts, $H = H_i + H^\prime$, where $H_i$ includes all parts associated with site i and $H^\prime$ represents the rest of the Hamiltonian.
A direct calculation leads to

\begin{equation}
<S_l^z S_i^z> = \Big< S_l^z\frac{ Tr_{(i)}S_i^z e^{-\beta H_i}}{Tr_{(i)}e^{-\beta H_i}}\Big> - \Big<S_l^z\Big[\frac{Tr_{(i)}S_i^z e^{-\beta H_i}}{Tr_{(i)}S_i^z e^{-\beta H_i}}-S_i^z\Big] \Delta \Big>\label{eq3}
\end{equation} 
where $Tr_{(i)}$ represents the partial trace with respect to site i and   $\Delta= 1 - e^{-\beta H_i}e^{-\beta H^\prime}e^{\beta (H_i+H^\prime)}$.
Equation (\ref{eq3}) is an exact relation. However, it is difficult to be used. Therefore, we will make an approximation based on the following decoupling,

\begin{equation}
\Big<S_l^z\Big[\frac{Tr_{(i)}S_i^z e^{-\beta H_i}}{Tr_{(i)}e^{-\beta H_i}}-S_i^z\Big]\Delta\Big> \approx \Big<S_l^z\Big[\frac{Tr_{(i)}S_i^z e^{-\beta H_i}}{Tr_{(i)}e^{-\beta H_i}}-S_i^z\Big]\Big> <\Delta>\label{eq4} 
\end{equation}  

Inserting (\ref{eq4}) into (\ref{eq3}) and using the fact that $<\Delta> \leq  1$, we obtain,
\begin{equation}
<S_l^zS_i^z> \leq  \Big<S_l^z\frac{ Tr_{(i)} S_i^z e^{-\beta H_i}}{Tr_{(i)} e^{-\beta H_i}}\Big>. \label{eq5}
\end{equation} 
By expanding $\Delta$ we see that the approximation is correct to the order of $\beta^{2}$. Moreover, it is consistent with the application of the correlation inequalities that will be used later to obtain the upper bounds for the critical couplings. In the next steps we will keep only the = sign of (\ref{eq5}) .

Let us write $H_i = -E_i S_i^z - \Omega  S_i^x$, where $E_i = \sum_j J_{ij}S_j^z$. Diagonalizing and taking the partial trace over i , we get for the longitudinal spin correlation function, $<\sigma_l^z \sigma_i^z>$, where $\sigma_i=2S_i$,
\begin{eqnarray}
&&<\sigma_l^z \sigma_i^z>= \Big<\sigma_l^z\frac{\sum_j J_{ij}\sigma_j^z}{\sqrt{(2\Omega)^2 + (\sum_j J_{ij}\sigma_j^z)^2}}\times \nonumber\\
&&\tanh\Big(\beta\sqrt{(2\Omega)^2 + (\sum_j J_{ij}\sigma_j^z)^2} \Big) \Big>.\label{eq6}
\end{eqnarray}

Introducing the exponential operator 
$e^{(aD )}.f(x) = f(x+a)$ , $D =\frac{\partial}{\partial x}$, 
we obtain,
\begin{eqnarray}
&&<\sigma_l^z \sigma_i^z>= \Big<\sigma _l^z e^{\sum_j (J_{ij}\sigma_j^z) D}\Big>\cdot f(x)\mid _{x=0}\nonumber\\
&&=\Big<\sigma _l^z\sigma_j^z  \prod_je^{(J_{ij}\sigma_j^z)D}\Big>\cdot f(x)\mid _{x=0}\label{eq7}
\end{eqnarray}  
where f(x) is given by
\begin{equation}
f(x) = \frac{x}{\sqrt{(2\Omega)^2 + x^2}}\tanh(\beta)\sqrt{(2\Omega)^2 + x^2} \label{eq8}
\end{equation} 
Note that $f(x)= -f(-x)$.

Expanding the exponential in (\ref{eq7}) and considering that $(\sigma_i^z)^2=1$, we obtain,
\begin{equation}
<\sigma_l^z \sigma_i^z>=\Big<\sigma_l^z\prod_j\Big[\cosh(J_{ij}D)+\sigma_i^z\sinh(J_{ij}D)\Big]\Big>\nonumber\\
\cdot f(x)\mid _{x=0}\label{eq9}
\end{equation} 
By a similar procedure the transverse two-spin correlation function $<$$\sigma_l^x \sigma_i^x$$>$ is obtained,
\begin{equation}
<\sigma_l^x \sigma_i^x>= \Big<\sigma_l^x\prod_j\Big[\cosh(J_{ij}D) +\sigma_i^z \sinh(J_{ij}D)\Big]\Big>\cdot g(x)\mid _{x=0} \label{eq10}
\end{equation}
where g(x) is
\begin{equation}
g(x) = \frac{2\Omega}{\sqrt{(2\Omega)^2 + x^2}}\tanh(\beta)\sqrt{(2\Omega)^2 + x^2} = g(-x)\label{eq11}
\end{equation} 
The expectation value of $\sigma_i^x$ is given by,
\begin{equation}
< \sigma_i^x>= \Big<\prod_j\Big[\cosh(J_{ij}D) +\sigma_i^z \sinh(J_{ij}D)\Big]\Big>\cdot g(x)\mid _{x=0}\label{eq12}
\end{equation} 

\section{Application to d=2 square lattice and d=3 cubic lattice.}
\subsection{d=2 lattice}
Considering the four neighbours of $i$ in (\ref{eq9}), expanding the product, applying the exponential operators appearing in the powers of $\cosh(J_{ij}D)$ and $\sinh(J_{ij}D)$ in f(x), we obtain,
\begin{equation}
<\sigma_l^z \sigma_i^z>= A_2\sum_j<\sigma_l^z \sigma_j^z>+B_2\sum_{j<k<m}<\sigma_l^z \sigma_j^z\sigma_k^z \sigma_m^z>\label{eq13}
\end{equation} 
where
\begin{eqnarray}
&&A_2 = \frac{1}{8} [f(4 J) + 2f(2 J)] >0 \nonumber\\
&&B_2 = \frac{1}{8}[f(4 J) - 2f(2 J)] <0 \nonumber\\ \label{eq14}
\end{eqnarray}
and $j$, $k$ and $m$ are neighbours of $i$ and $f(..)$ is given by (\ref{eq8}).
\subsection{d=3 lattice.}
After a similar calculation we obtain for the cubic lattice,
\begin{eqnarray}
&&<\sigma_l^z \sigma_i^z>= A_3\sum_j<\sigma_l^z \sigma_j^z> +B_3\sum_{j<k<m}<\sigma_l^z \sigma_j^z\sigma_k^z \sigma_m^z>  \nonumber\\
&&+C_3\sum_{j<k<m<n<p} \!\!\!\!<\!\!\sigma_l^z \sigma_j^z\sigma_k^z \sigma_m^z \sigma_n^z \sigma_p^z>\label{eq15}
\end{eqnarray} 
where,
\begin{eqnarray}
&&A_3 = \frac{1}{2^5} [f(6 J) + 4f(4 J) + 5f(2 J)] >0 \nonumber\\
&&B_3 = \frac{1}{2^5}[f(6 J) - 3f(2 J)] <0 \nonumber\\
&&C_3 = \frac{1}{2^5}[f(6 J) - 4f(4 J) + 5f(2 J)] >0 \nonumber\\\label{eq16}
\end{eqnarray} 

and $j$, $k$, $m$, $n$ and $p$ are neighbours of $i$ and $f(..)$ is given by (\ref{eq8}).

\section{Exponential decay of the two-point functions and the upper bounds.}

Upper bounds for the critical temperature $T_c$ for Ising and multi-component spin systems have been obtained by showing (for $T$$>$$T_c$) the exponential decay of the two-point function \cite{Fisher1967,Simon1980,Brydges1982}. The procedure to obtain these upper bounds for the critical couplings of the tranverse Ising model is the following: we start from a two-point correlation function equations (\ref{eq13}) and (\ref{eq15})  and we make use of Griffiths  inequalities (Griffiths I, II) \cite{Gallavotti1971,Suzuki1973,Contucci2010} and Newman and Lebowitz inequalities \cite{Newman1975,Contucci2010} . A proof of Griffiths inequalities has been given for the XY model with no external field \cite{Gallavotti1971}. Extensions of Griffiths-Kelly-Sherman inequalities to quantal systems, under external fields, both longitudinal and transverse, have been proved \cite{Suzuki1973,Contucci2010}. 
The physical reason why the Griffiths and similar inequalities are valid for the quantal XY-type Hamiltonian is that the off-diagonal interaction, namely, $
 H_1(x)=\sum_{A} J_{A}^x\sigma_A^x, (J_{A}^x\leq0)$, produce the decrease of the ferromagnetic correlation among the $\sigma_j^z $-spins, but it is not sufficient big to create a cooperative effect to induce an antiferromagnetic correlation. In other words, one can say that $H_1(x)$ is a dynamical random force acting on z-z correlations \cite{Suzuki1973}.
We establish the inequality for the two-point function $<$$\sigma_l^z \sigma_i^z$$>$,

\begin{equation}
        <\sigma_l^z \sigma_i^z>\leq   \sum_ja_j <\sigma_l^z \sigma_j^z>, 0\leq a_j\leq 1,   \label{eq17}
\end{equation} 
which when iterated \cite{Simon1980} implies exponential decay for $T>T_c$.

\subsection{Upper bounds for d=2.}

From Eq.(\ref{eq13}), using Griffiths II ($<$$\sigma_l^z \sigma_j^z\sigma_k^z \sigma_m^z$$>\geq<$$\sigma_l^z \sigma_j^z$$><$$\sigma_k^z \sigma_m^z$$>$) in the second term and considering $B_2$$<$$0$, we get,
\begin{equation}
<\sigma_l^z \sigma_i^z>\leq  \sum_ja_j<\sigma_l^z \sigma_j^z>\label{eq18}, 
\end{equation} 
where $j \neq k$ are neighbours of $i$,
and 
\begin{equation}
a_j=A_2-|B_2|<\sigma_j^z \sigma_k^z>_{1d}\label{eq19}
 \end{equation}

\subsection{ Upper bounds for d=3.}

From Eq.(\ref{eq15}), using Griffiths II in the second term $B_3$$<$$0$, Newman's inequality $(<$$\sigma_i^zF$$>\leq\sum_{j}<$$\sigma_i^z\sigma_j^z$$><$$dF/d\sigma_j^z$$>$, F are polynomials with positive coefficients) combined with Griffiths I $(<$$\sigma_A^z$$>\leq1)$  on the third term $C_3$$>$$0$ , we get,
\begin{equation}
<\sigma_l^z \sigma_i^z>\leq  \sum_j a_j<\sigma_l^z \sigma_j^z>\label{eq20}, 
\end{equation} 
where $j\neq k$ are neighbours of $i$,
and
\begin{equation} 
a_j=A_3-|B_3|<\sigma_j^z \sigma_k^z>_{1d} + 5 C_3 \label{eq21}
\end{equation}
\subsection{Numerical Results}
The two-spin correlation functions appearing in (\ref{eq19}) and (\ref{eq21}) $<\sigma_j^z \sigma_k^z>_{1d}$  is the one-dimensional model two-spin correlation function separated by a distance of two lattices sites. For the one-dimensional transverse Ising model the exact value of this function  at the critical value $\Omega_c  = J$ is \cite{Pfeuty1970}:
\begin{equation}
<\sigma_j^z \sigma_k^z>_{1d} = \frac{1}{4}\Big(\frac{2}{\pi} \Big)^2 8 \frac{H^{4}(2)}{H(4)} \label{eq22}
\end{equation}
where $H(n)= 1^{n-1} 2^{n-2}...(n-1)$. 
For the one-dimensional Ising model the exact value of the  spin correlation function separated by a distance of two lattices sites is:
\begin{equation}
<\sigma_j^z \sigma_k^z>_{1d}=\tanh^{2}\beta J \label{eq23}
\end{equation}

\begin{figure}[b]
\begin{center}
\includegraphics[angle=0,width=0.85\textwidth]{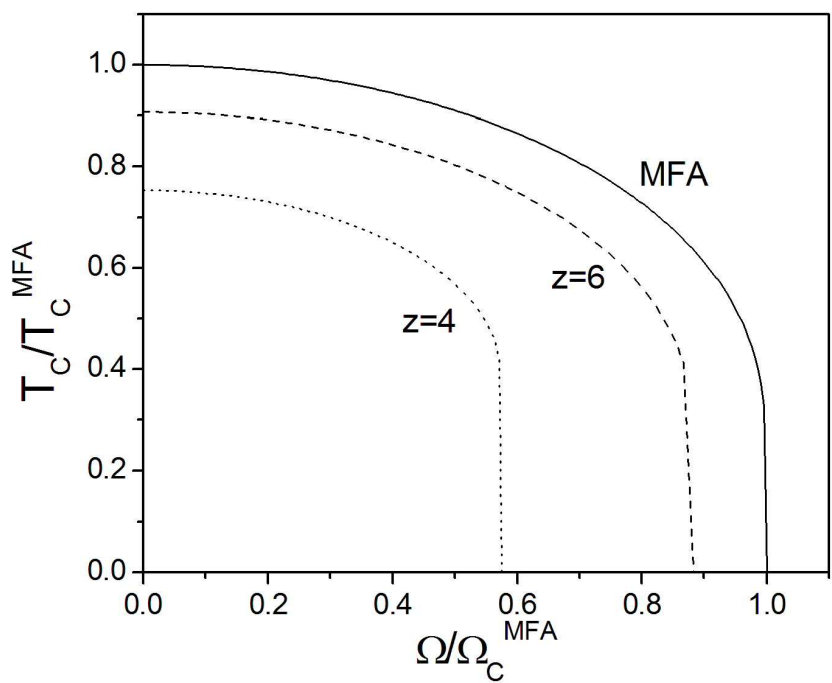}
\end{center}
\caption{Phase diagram of the TIM model for $z=4$ and $z=6$. Solid line is the mean field approximation result. Results for our approach evaluated by using the spin correlation function separated by two sites, Eq.(\ref{eq23}) are given by the dotted ($z=4$) and dashed ($z=6$) lines.}
\label{figure1}
\end{figure}

In the following, we will use result (\ref{eq22}) in (\ref{eq19}) and (\ref{eq21}) to calculate the upper bound for $\Omega_c$ at $T_c=0$ and result (\ref{eq23}) in (\ref{eq19}) and (\ref{eq21}) to calculate the upper bound for  $T_c$ at $\Omega=0$. 
Evaluating numerically the value of T such that $\sum_j  a_j \leq 1$, $a_j > 0$, we obtain, by sufficient  condition (see Eq.(\ref{eq17})) , the upper bounds for $T_c$ as a function of $\Omega$, shown in figure \ref{figure1}, together with the curve for the mean field results. We use (\ref{eq23}) for the one-dimensional two-spin correlation function in obtaining figure \ref{figure1}. This curve represents the rigorous upper bounds for  ${T_c, \Omega_c}$. In particular, the mean field values are $\Omega_c^{MFA} = zJ/2$ and $T_c^{MFA}(0) = zJ/k_B$. The values for $\Omega_c$ and $T_c(0)$, which are the rigorous upper bounds for $d=2,z=4$ and $d=3, z=6$, obtained in the present calculation are:
(a)	d=2, z=4; $\frac{k_BT_c}{J} =3.014$ and $\Omega_c = 1.3755 J$,
(b)	d=3, z=6; $\frac{k_BT_c}{J} = 5.423$  and $\Omega_c = 2.4466J$.

In table \ref{tab}, we compare the results obtained by the effective field calculation (EFT) \cite{SaBarreto1981}, the high temperature expansion (HTE) \cite{Elliott1971,Pfeuty1971} and the present results for $\Omega_c$.
\begin{table}
\centering
\caption{Estimatives for $\Omega_c/\Omega_cˆ{MFA}$ for d=2 and d=3.} 
 
\begin{tabular}{ccc} %

\hline
        &   $d=2, z=4$   &   $d=3, z=6$   \\
\hline
MFA & 1 &1 \\
EFT & 0.688 & 0.784 \\
HTE & 0.770 & 0.860 \\
Present work & 0.643 & 0.813 \\
\hline
\end{tabular}
\label{tab}
\end{table}

\section{Concluding remarks}
In this paper we have obtained rigorous upper bounds for the critical couplings of the transverse Ising model. The procedure was based on an approximation for an exact identity for the two-spin correlation functions and on rigorous inequalities for the spin correlation functions. The approximated relation for the two-spin correlation function, Eq.(\ref{eq5}), used in this procedure, is consistent with the rigorous inequalities, Eq.(\ref{eq17}), since both act in the same direction of the inequalities. The upper bounds were applied for two- and three- dimensional models.

\section*{Acknowledgements}FCSB is grateful for the financial support of CAPES/Brazil which made possible his visit to the UFSJ/Brasil. ALM acknowledges financial support from CNPq/Brazil and FAPEMIG/Brazil.

\vspace{1cm}

\end{document}